# Theoretical study of the thermoelectric properties of SiGe nanotubes

J. Wei[1], H. J. Liu[1,*], X. J. Tan[1], L. Cheng[1], J. Zhang[1], D. D. Fan[1], J. Shi[1], X. F. Tang[2]

[1]*Key Laboratory of Artificial Micro- and Nano-structures of Ministry of Education and School of Physics and Technology, Wuhan University, Wuhan 430072, China*

[2]*State Key Laboratory of Advanced Technology for Materials Synthesis and Processing, Wuhan University of Technology, Wuhan 430070, China*

The thermoelectric properties of two typical SiGe nanotubes are investigated using a combination of density functional theory, Boltzmann transport theory, and molecular dynamics simulations. Unlike carbon nanotubes, these SiGe nanotubes tend to have gear-like geometry, and both the (6, 6) and (10, 0) tubes are semiconducting with direct band gaps. The calculated Seebeck coefficients as well as the relaxation time of these SiGe nanotubes are significantly larger than those of bulk thermoelectric materials. Together with smaller lattice thermal conductivity caused by phonon boundary and alloy scattering, these SiGe nanotubes can exhibit very good thermoelectric performance. Moreover, there are strong chirality and temperature dependence of the *ZT* values, which can be optimized to 4.9 at room temperature and further enhanced to 5.4 at 400 K for the armchair (6, 6) tube.

## 1. Introduction

Due to increasing energy crisis and environment pollution, there is currently growing interests in searching for advanced energy materials. Among them, the thermoelectric materials which can directly convert heat into electricity or vice versa have attracted much attention. The performance of a thermoelectric material at temperature $T$ can be characterized by the dimensionless figure of merit:

$$ZT = S^2 \sigma T / (\kappa_e + \kappa_p), \qquad (1)$$

---

[*] Author to whom correspondence should be addressed. Electronic mail: phlhj@whu.edu.cn

which includes the Seebeck coefficient $S$, the electrical conductivity $\sigma$, the electronic thermal conductivity $\kappa_e$, and the phonon derived thermal conductivity $\kappa_p$. In order to compete with the traditional energy conversion methods, the ZT value of a thermoelectric material should reach at least 3. In principle, the ZT value can be improved by utilizing some strategies to increase the power factor ($S^2\sigma$) and/or decrease the thermal conductivity ($\kappa_e + \kappa_p$). However, it is usually challenging to do so because of strong interdependence of these transport coefficients, and the ZT values of most good thermoelectric materials reported so far are found to be 1~2. Among them, silicon germanium (SiGe) alloys have long been used in thermoelectric modules for space missions to convert radio-isotope heat into electricity [1]. During the past 50 years, many efforts have been made to improve the thermoelectric performance of SiGe alloys. It was found that the highest ZT value of about 1.5 can be reached at 900°C for n-type $Si_{0.8}Ge_{0.2}$ alloy [2]. For the p-type systems, however, the ZT values are relatively lower and a peak value of 0.95 can be obtained at temperature of 800~900°C [3]. In the early 1990s, Hicks and Dresselhaus [4] theoretically predicated that one- and two-dimensional structures could have significantly larger ZT values than the corresponding bulk materials. The reason is that low-dimensional systems can induce enhanced phonon boundary scattering so that the thermal conductivity is reduced. Moreover, the power factor may be increased by quantum confinement and energy filtering effects. Inspired by such concept, a lot of subsequent works have been devoted to the fabrications of various low-dimensional thermoelectric materials. For example, the $Bi_2Te_3/Sb_2Te_3$ superlattice thin film [5], the PbSeTe/PbTe quantum dot superlattice [6], and the Si nanowires [7, 8] have been successfully fabricated and their enhanced thermoelectric performances are confirmed. In the case of SiGe systems, Martinez *et al.* reported that p-type SiGe nanowires [9] were epitaxially grown on a Si (111) substrate, and the ZT value is increased by more than a factor of two over the single-crystal alloys [10]. Using the vapor-liquid-solid (VLS) method, Lee *et al.* [11] fabricated the n-type SiGe nanowire and found the ZT value is twice that of the radioisotope thermoelectric generator (RTG) sample [1].

Recently, the SiGe thin films [12] were successfully grown using the electrophoresis deposition technique and the measured power factor at 950°C is larger than that of bulk alloy, suggesting the favorable thermoelectric performance. On the theoretical side, the nonequilibrium molecular dynamics (NEMD) simulations and Boltzmann theory predicted that the *ZT* value of *n*-type SiGe nanowires can be enhanced to 2.2 at 800 K [11]. By using a combination of density function theory (DFT) and nonequilibrium Green's function (NEGF) methods, Shi *et al.* [13] found that the maximum *ZT* values for *n*-type and *p*-type SiGe nanowires can be reached to 4.7 and 2.7, respectively. All these works suggest that low-dimensional SiGe can indeed significantly enhance the thermoelectric performance, which is mainly attributed to the reduction of lattice thermal conductivity caused by phonon boundary scattering.

Except for the SiGe thin films and nanowires, another important low-dimensional form of SiGe family is the nanotube structures, which can be viewed as rolling SiGe sheet into a cylindrical shape. Unlike carbon nanotube, the SiGe nanotube exhibit a gear-like geometry since the Si and Ge atoms tend to have $sp^3$ rather than $sp^2$ hybridization. Indeed, the molecular dynamics simulations by Zang *et al.* [14] found that a bilayer SiGe nanofilm may bend into a nanotube with Ge as the inner layer. The existence of stable single walled tubular form of SiGe has been justified by Rathi *et al.* [15] using hybrid DFT and finite cluster approach. They found that such kinds of SiGe nanotubes are semiconducting in nature, with a wide spectrum of band gaps. Liu *et al.* [16] investigated the structure and energetics of a series of SiGe nanotubes using *ab-initio* method and classical molecular dynamics simulations. Their calculated results indicate that large diameter nanotubes are more stable than small ones. By using first-principles pseudopotential method, Pan *et al.* [17] discussed the chirality and diameter dependence of the energetics and electronic properties of gear-like SiGe nanotubes. It should be mentioned that most of these works are focused on the structural and electronic properties of SiGe nanotubes, while their thermoelectric properties are less known so far, and it is thus the subject of the present work. We will consider two typical SiGe nanotubes (10, 0) and (6, 6), which have similar diameters but quite different chiralities. We shall see that by appropriately optimizing the carrier concentration and operation temperature, a highest *ZT* value of 1.8 and 5.4 can be achieved for the (10, 0) and (6, 6) tubes, respectively.

## 2. Computational Method

The structure optimization and electronic properties of SiGe nanotubes are calculated by using a plane-wave pseudopotential formulation [18, 19, 20] within the framework of DFT. The code is implemented in the Vienna *ab initio* simulation package (VASP). The exchange-correlation energy is in the form of Perdew-Wang-91 (PW91) [21] with generalized gradient approximation (GGA). The cutoff energy for the plane-wave expansion is taken to be 400 eV. The **k** points are sampled on a uniform grid along the tube axis. We adopt a supercell geometry and the tubes are aligned in a hexagonal array. The closest distance between each nanotube and its periodic image is 17 Å so that they can be treated as independent entities. The system is fully relaxed until the magnitude of the forces acting on all the atoms becomes less than 0.01 eV/Å. The electronic transport coefficients are derived by using the semi-classical Boltzmann theory [22], where the relaxation time is estimated from the deformation potential (DP) theory proposed by Bardeen and Shockley [23]. For the phonon transport, the lattice thermal conductivity is predicted using equilibrium molecular dynamics (EMD) simulations combined with the Green-Kubo autocorrelation decay method [24]. We use Tersoff potential [25] to describe the interatomic interactions, and the time step is set to 0.5 fs. After a constant temperature simulation of 500,000 steps and a constant energy simulation of 400,000 steps, the heat current data are collected to calculate the heat current autocorrelation function.

## 3. Results and Discussions

Figure 1 shows the ball-and-stick models of armchair (6, 6) and zigzag (10, 0) SiGe nanotubes. As mentioned above, the Si and Ge atoms prefer the $sp^3$ hybridization, and the side-view of SiGe nanotube looks gear-like. In principle, each SiGe nanotube can have two atomic configurations. For the type I, all the Si atoms form the inner shell and the Ge atoms form the outer one. It is just reversed for the type II. Here we only consider type I since previous work [17] indicates that at relatively large diameters, the energies and electronic properties of these two types are essentially the same. Note the diameter of SiGe nanotube is averaged between that of the inner and outer shell, which is calculated to be 6.54 Å for the (6, 6) tube and 6.29 Å for the (10, 0) tube. The buckling distance of these two tubes are respectively 0.64 Å and 0.66 Å, which are larger than that of SiGe monolayer due to the curvature effect [26]. Here, the similar diameter of (6, 6) and (10, 0) tubes offers us a good opportunity to study the chirality dependence of their electronic and transport properties. In Figure 2 we present the calculated energy band structures for these two kinds of tubes. It is well known that carbon nanotube with indices (n, m) is metallic if n-m is an integer multiple of 3, otherwise it is semiconducting. However, this is not necessarily the case for our SiGe nanotubes. We see from Fig. 2 that both (10, 0) and (6, 6) tubes are semiconducting with direct band gaps. In the case of (6, 6) tube, both the valence band maximum (VBM) and the conduction band minimum (CBM) appear at about 2/3 ΓX away from the Γ point, and the DFT calculated band gap is 0.40 eV. For the (10, 0) tube, the VBM and CBM are located at Γ point and the band gap is calculated to be 0.31 eV.

On the basis of the computed band structures, we are able to evaluate the transport coefficients by using the semi-classical Boltzmann theory [22] and the rigid-band approach [27]. Within this method, the Seebeck coefficient $S$ is independent of the relaxation time $\tau$, while the electrical conductivity $\sigma$ and the power factor $S^2\sigma$ can only be calculated with respect to $\tau$. Here, the relaxation time is obtained by applying the DP theory [23] combined with the effective mass approximation. For one-dimensional (1D) systems, the relaxation time can be expressed as [28]:

$$\tau = \frac{\hbar^2 C}{(2\pi k_B T)^{1/2} \left|m^*\right|^{1/2} E_1^2}, \qquad (2)$$

where $m^*$ is the effective mass, $C$ is the elastic constant, and $E_1$ is the deformation potential constant. These three quantities can be readily obtained from first-principles calculations. The other parameters $\hbar$, $k_B$ and $T$ are the reduced Planck constant, the Boltzmann constant, and the absolute temperature, respectively. The calculated relaxation time at room temperature is summarized in Table I. For the (6, 6) tube, it is found that the effective mass of electron and hole have similar values while the deformation potential constant of electron is more than twice larger than that of hole. As a result, the relaxation time of hole is significantly larger than that of electron. However, this is not the case for the (10, 0) tube, where the electron has a relatively larger relaxation time due to smaller deformation potential constant. For both tubes, the relaxation time is one or two orders of magnitude larger than those found in most thermoelectric materials, which suggests that SiGe nanotubes may have better thermoelectric performance. Moreover, if we compare (6, 6) with (10, 0) tube, we see the former has much larger relaxation time for both electron and hole, which makes it more favorable for the thermoelectric applications. We will come back to this point later. In Figure 3, we plot the calculated relaxation time as a function of temperature that ranges from 300 K to 1000 K. As the electron scattering is more frequent at high temperatures, we find that the relaxation times of both (6, 6) and (10, 0) tubes decrease with increasing temperature and roughly follow an exponential decay law. In the whole temperature range, the relaxation times of (6, 6) tube are always larger than those of (10, 0) tube, which indicates the important role of tube chirality.

With the relaxation time obtained, all the electronic transport coefficients can be determined [29, 30]. Figure 4(a) shows the calculated Seebeck coefficients $S$ as a function of chemical potential $\mu$ at 300 K. Within the rigid-band picture [27], the chemical potential indicates the doping level (or carrier concentration) of the system. For *n*-type doping the chemical potential is positive $(\mu > 0)$ while it is negative $(\mu < 0)$ for *p*-type doping. It can be seen that the Seebeck coefficients of SiGe nanotubes exhibit two obvious peaks around the Fermi level $(\mu = 0)$, and the peak values of (6, 6) tube are slightly larger than those of (10, 0) tube. This is reasonable since the former has a larger band gap. As for the electrical conductivity $\sigma$, we see from Figure 4(b) there is a sharp increase of $\sigma$ around the band edge which is more pronounced for the (6, 6) tube. This is consistent with the fact that (6, 6) tube have much larger relaxation time compared with (10, 0) tube. Comparing Figure 4(a) with 4(b), we find that at the chemical potential where the Seebeck coefficient reaches the peak value, the electrical conductivity is actually very small. On the other hand, at the chemical potential when there is a sharp increase of the electrical conductivity, the Seebeck coefficient becomes very small. This contradictory behavior suggests that there must be a trade-off between the Seebeck coefficient $S$ and the electrical conductivity σ such that the power factor $(S^2\sigma)$ can be maximized at a particular doping level or carrier concentration. Indeed, we see from Figure 4(c) that by appropriate doping, the power factor of SiGe nanotubes can be enhanced to larger values. For the (6, 6) tube, the optimized carrier concentration has a chemical potential $\mu$ = 0.19 eV for *n*-type and $\mu$ = −0.18 eV for *p*-type. The corresponding values for the (10, 0) tube are $\mu$ = 0.12 eV and $\mu$ = −0.18 eV, respectively. Note that the power factor of (6, 6) tube is much larger than that of the (10, 0) tube, especially for the *p*-type doping. Once again, this suggests that the armchair (6, 6) may be more favorable than the zigzag (10, 0) when using SiGe nanotubes as possible thermoelectric materials. On the other hand, the electronic thermal conductivity $\kappa_e$ plotted in Figure 4(d) shows a similar behavior as the electrical conductivity. This is reasonable since $\kappa_e$ is derived from the $\sigma$ by using the Wiedemann-Franz law:

$$\kappa_e = L\sigma T, \qquad (3)$$

where a constant Lorentz number $L = 2.44 \times 10^{-8} V^2/K^2$ is used for the SiGe systems [9, 11].

We next discuss the phonon transport of these SiGe nanotubes by using the EMD method, where the phonon-derived thermal conductivity $\kappa_p$ can be expressed as an integration of heat current autocorrelation:

$$\kappa_p = \frac{1}{Vk_B T^2} \int_0^{\tau_m} \langle J(\tau)J(0)\rangle d\tau . \qquad (4)$$

Here $V$ is the system volume, $k_B$ is the Boltzmann constant, $T$ is the system temperature, $J$ is the heat current, $\tau_m$ is the integration time, and the angular brackets denote an average over time in the MD simulation. The advantage of the EMD method is that it does not require an imposed driving force which may have influence on the heat flux and the thermal conductivity. However, it has been established that finite-size effects do play a role in applying the EMD method [31, 32, 33]. We thus perform convergence test and find that a supercell containing 2400 and 2000 atoms are needed to obtain reliable results for the (6, 6) and (10, 0) tubes, respectively. Figure 5 plots the calculated thermal conductivity $\kappa_p$ of these two nanotubes as a function of temperature. Note the definition of the cross-sectional area has some arbitrariness for low-dimensional systems such as our SiGe nanotubes. To be consistent with the calculations of electronic transport coefficients discussed above, here the value of lattice thermal conductivity is recalculated with respect to a hexagonal cell having dimensions of 30 Å × 30 Å. We see that for the (6, 6) tube, $\kappa_p$ decreases exponentially with increasing temperature and can be fitted as:

$$\kappa_p = 2.05 + 11.36 \times e^{-T/176.54} . \qquad (5)$$

The case for the (10, 0) tube is similar, which almost coincides with that of (6, 6) tube at intermediate temperature region. However, at lower and higher temperature region, the thermal conductivity of (10, 0) tube exhibits a slightly lower value. The temperature dependence of $\kappa_p$ can be also fitted as:

$$\kappa_p = 1.72 + 3.61 \times e^{-T/366.04} . \qquad (6)$$

Note the thermal conductivity of these two nanotubes are significantly smaller than that of bulk Si (156 W/mK at 300 K [34], and 65 W/mK at 1000 K [31]). Even we use a "realistic" cross-sectional area defined by $\pi r^2$ (where $r$ is the radius of tube), the calculated thermal conductivity of SiGe nanotubes are still much lower than the bulk values, which is believed to be caused by a combination of alloy scattering and boundary scattering [35, 36].

Inserting all the transport coefficients into Eq. (1), we are now able to evaluate the thermoelectric performance of SiGe nanotubes. Figure 6 plots the room temperature *ZT* value as a function of chemical potential for the (6, 6) and (10, 0) tubes. We see there are two remarkable peaks around the Fermi level which corresponds to *p*-type and *n*-type doping. For the (6, 6) tube, the *ZT* value can be optimized to 4.9 at $\mu = -0.10$ eV, and 2.3 at $\mu = 0.14$ eV. Such *ZT* values not only significantly exceed that of bulk SiGe, but are also higher than those of other low-dimensional SiGe systems reported so far. On the other hand, the calculated *ZT* values of (10, 0) tube are obviously smaller than those of (6, 6) tube, which is around 1.0 for both *p*-type and *n*-type doping. Detailed analysis of the transport coefficients find that (6, 6) tube actually has larger electronic and lattice thermal conductivity compared with (10, 0) tube, the better thermoelectric performance of the former can be attributed to its much larger power factor (especially the Seebeck coefficient) at optimized carrier concentration. Our calculations suggest that the chirality of nanotubes plays an important role in determining their thermoelectric performance.

In Figure 7, we plot the calculated *ZT* value of these two SiGe tubes as a function of temperature at optimal chemical potential. For the (10, 0) tube, the temperature dependence of *ZT* value is weak, which varies from 0.9~1.8 in the whole temperature range from 300 K to 1000 K. In the case of (6, 6) tube, however, we see that the *ZT* values of both *n*-type and *p*-type increase with temperature, and reach a peak at 400 K, followed by a slow decay. The maximum *ZT* value can be as high as 5.4 for *p*-type doping and 3.1 for *n*-type doping. It should be mentioned that in a wide temperature range from 300 K to about 700 K, the *ZT* values of (6, 6) tube are always larger than 2.0. All these findings make SiGe nanotube a very promising candidate for thermoelectric applications. Our calculated *ZT* values are summarized in Table II where the corresponding chemical potential and transport coefficients are also given.

## 4. Summary

In summary, our theoretical calculations demonstrate that SiGe nanotubes could be optimized to exhibit very good thermoelectric performance. The predicted *ZT* value of armchair (6,6) is larger than that of the zigzag (10,0), and a maximum *ZT* value of 5.4 can be achieved at 400 K, which not only exceeds those of the best bulk thermoelectric materials, but are also competitive to those of other low-dimensional systems reported so far. It should be mentioned that SiGe nanotubes with a minimum diameter of 50 nm have been fabricated by rolling thin solid SiGe films [37]. Moreover, the existence of stable single walled SiGe nanotubes has been justified by theoretical studies [14, 15]. It is thus reasonable to expect that if the number of walls, the diameter, and the chirality can be experimentally controlled, SiGe nanotubes could become very promising thermoelectric materials.

## Acknowledgements

This work was supported by the National Natural Science Foundation (Grant No. 51172167 and J1210061) and the "973 Program" of China (Grant No. 2013CB632502).

**Table I** The relaxation time $\tau$ at 300 K for the (6, 6) and (10, 0) SiGe nanotubes. The corresponding carriers type, effective mass $m^*$, elastic constant $C$, and deformation potential constant $E_1$ are also given.

| System | Carriers | $m^*$ ($m_0$) | $C$ (eV/Å) | $E_1$ (eV) | $\tau$ (ps) |
|---|---|---|---|---|---|
| (6, 6) | electron | 0.134 | 131.12 | −1.33 | 0.91 |
|  | hole | 0.132 | 131.12 | −0.60 | 4.52 |
| (10, 0) | electron | 0.611 | 122.70 | −1.34 | 0.40 |
|  | hole | 0.432 | 122.70 | −2.08 | 0.19 |

**Table II** Optimized ZT values of the (6, 6) and (10, 0) SiGe nanotubes at different temperature. The corresponding chemical potential and transport coefficients are also indicated.

| System | $T$ (K) | $\mu$ (eV) | $S$ ($\mu$V/K) | $\sigma$ ($10^5$ S/m) | $S^2\sigma$ ($10^{-2}$W/mK$^2$) | $\kappa_e$ (W/mK) | $\kappa_p$ (W/mK) | ZT |
|---|---|---|---|---|---|---|---|---|
| (6, 6) nanotube | 300 | −0.10 | 448.00 | 8.28 | 16.63 | 6.09 | 4.17 | 4.9 |
|  |  | 0.14 | −353.62 | 4.81 | 6.02 | 3.54 |  | 2.3 |
|  | 400 | −0.07 | 419.19 | 9.48 | 16.65 | 9.29 | 3.05 | 5.4 |
|  |  | 0.11 | −361.40 | 4.17 | 5.45 | 4.09 |  | 3.1 |
|  | 500 | −0.07 | 344.30 | 16.44 | 19.49 | 20.14 | 2.94 | 4.2 |
|  |  | 0.10 | −331.80 | 5.22 | 5.74 | 6.39 |  | 3.1 |
|  | 600 | −0.07 | 281.78 | 29.93 | 23.77 | 44.00 | 2.36 | 3.1 |
|  |  | 0.09 | −293.38 | 6.99 | 6.01 | 10.27 |  | 2.9 |
|  | 700 | −0.09 | 237.53 | 49.46 | 27.90 | 84.82 | 2.29 | 2.2 |
|  |  | 0.10 | −259.43 | 52.24 | 35.16 | 89.59 |  | 2.7 |
|  | 800 | −0.10 | 205.17 | 70.61 | 29.72 | 138.40 | 2.17 | 1.7 |
|  |  | 0.11 | −236.37 | 15.58 | 8.71 | 30.54 |  | 2.1 |
|  | 900 | −0.11 | 181.15 | 91.78 | 30.12 | 202.38 | 2.12 | 1.3 |
|  |  | 0.12 | −221.39 | 19.77 | 9.69 | 43.59 |  | 1.9 |
|  | 1000 | −0.13 | 163.24 | 117.98 | 31.44 | 289.06 | 2.07 | 1.1 |
|  |  | 0.13 | −210.38 | 25.46 | 11.27 | 62.39 |  | 1.7 |
| (10, 0) nanotube | 300 | −0.14 | 281.72 | 2.44 | 1.93 | 1.79 | 3.32 | 1.1 |
|  |  | 0.09 | −270.54 | 2.09 | 1.53 | 1.53 |  | 0.9 |
|  | 400 | −0.12 | 299.50 | 2.12 | 1.90 | 2.08 | 2.96 | 1.5 |
|  |  | 0.08 | −265.69 | 2.08 | 1.47 | 2.04 |  | 1.2 |
|  | 500 | −0.11 | 291.46 | 2.28 | 1.94 | 2.80 | 2.55 | 1.8 |
|  |  | 0.08 | −233.38 | 2.78 | 1.51 | 3.40 |  | 1.2 |
|  | 600 | −0.10 | 269.55 | 2.74 | 1.99 | 4.03 | 2.49 | 1.8 |
|  |  | 0.09 | −203.34 | 4.20 | 1.74 | 6.18 |  | 1.2 |
|  | 700 | −0.10 | 243.86 | 3.45 | 2.05 | 5.92 | 2.27 | 1.8 |
|  |  | 0.11 | −187.89 | 6.15 | 2.17 | 10.55 |  | 1.2 |
|  | 800 | −0.10 | 222.04 | 4.12 | 2.03 | 8.08 | 2.07 | 1.6 |
|  |  | 0.13 | −183.00 | 8.33 | 2.79 | 16.33 |  | 1.2 |
|  | 900 | −0.11 | 202.23 | 5.06 | 2.07 | 11.16 | 1.92 | 1.4 |
|  |  | 0.16 | −182.03 | 10.80 | 3.58 | 23.81 |  | 1.2 |
|  | 1000 | −0.12 | 186.18 | 5.95 | 2.06 | 14.58 | 1.66 | 1.3 |
|  |  | 0.18 | −181.07 | 13.60 | 4.46 | 33.33 |  | 1.3 |

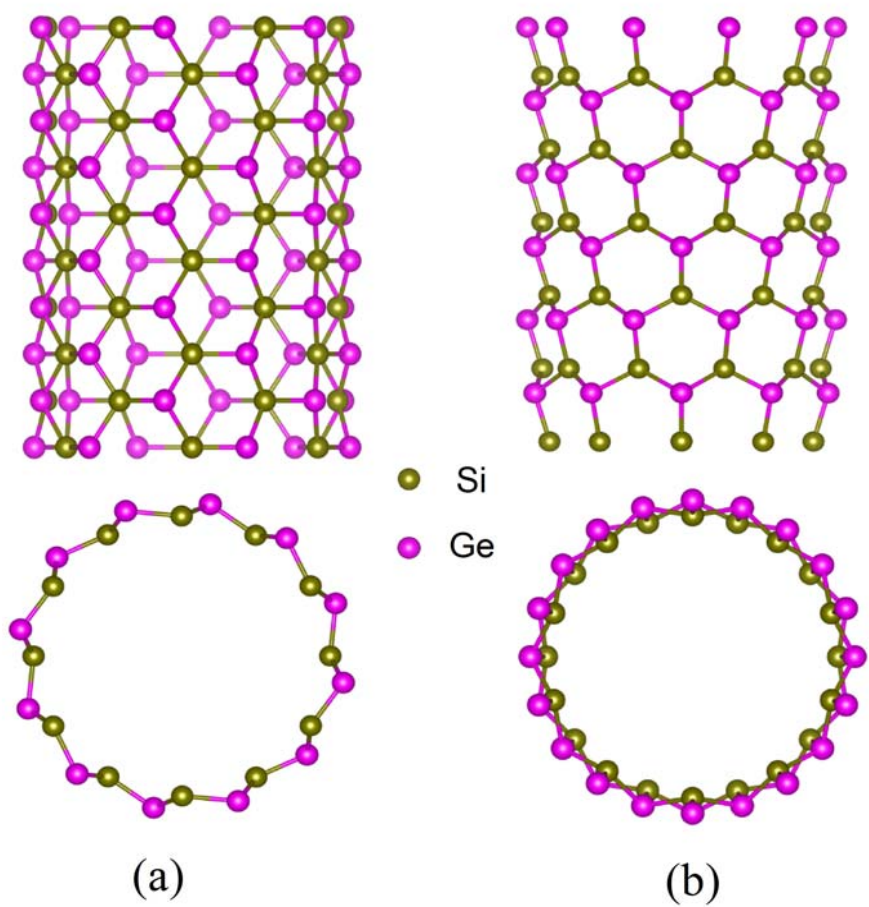

**Fig. 1** Top- and side- views of the optimized structures for the SiGe nanotubes: (a) (6, 6), and (b) (10, 0).

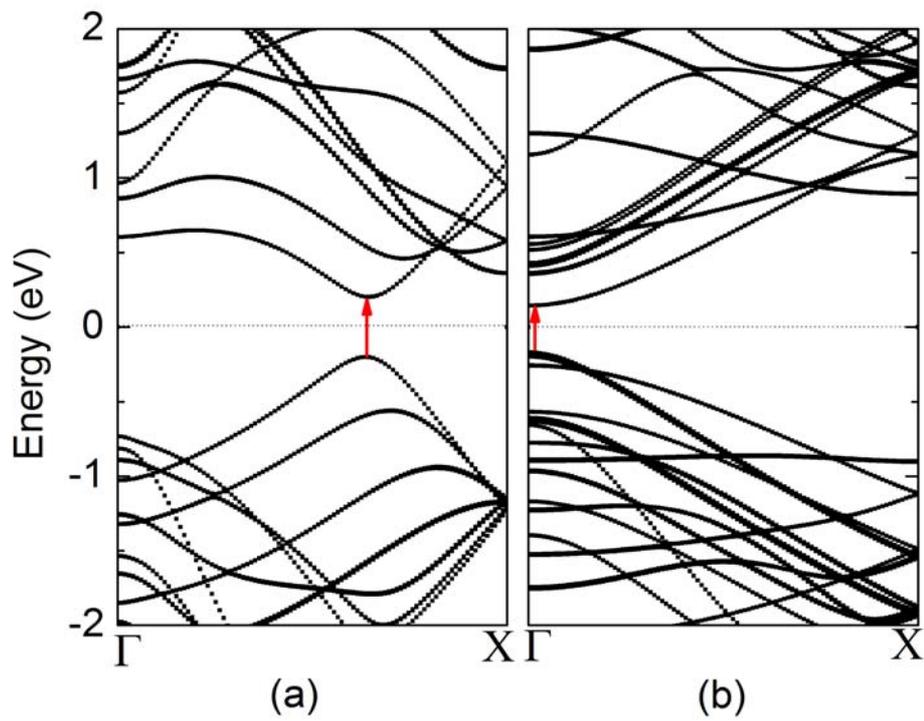

**Fig. 2** Energy band structures of the SiGe nanotubes: (a) (6, 6), and (b) (10, 0). The Fermi level is at 0 eV.

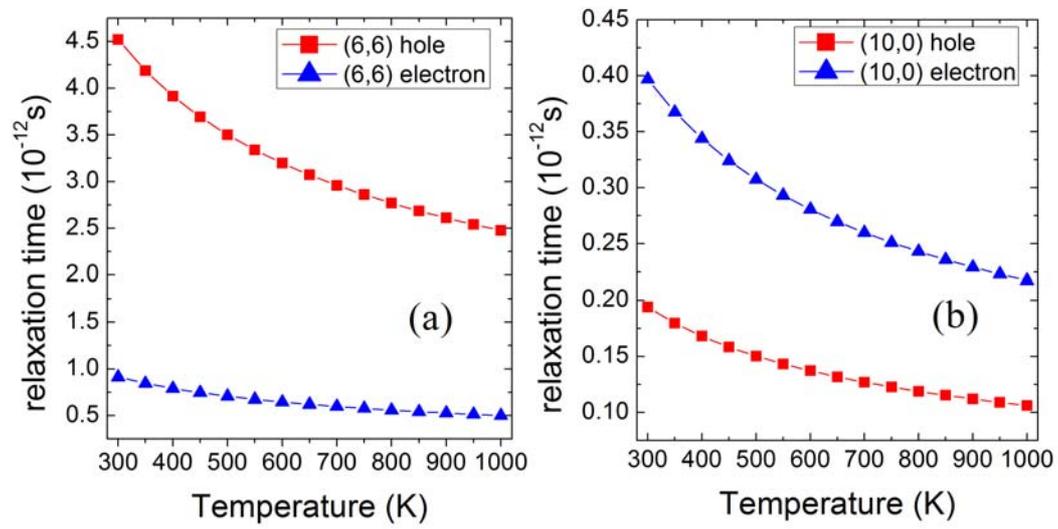

**Fig. 3** Calculated carriers relaxation time as a function of temperature for the SiGe nanotubes: (a) (6, 6), and (b) (10, 0).

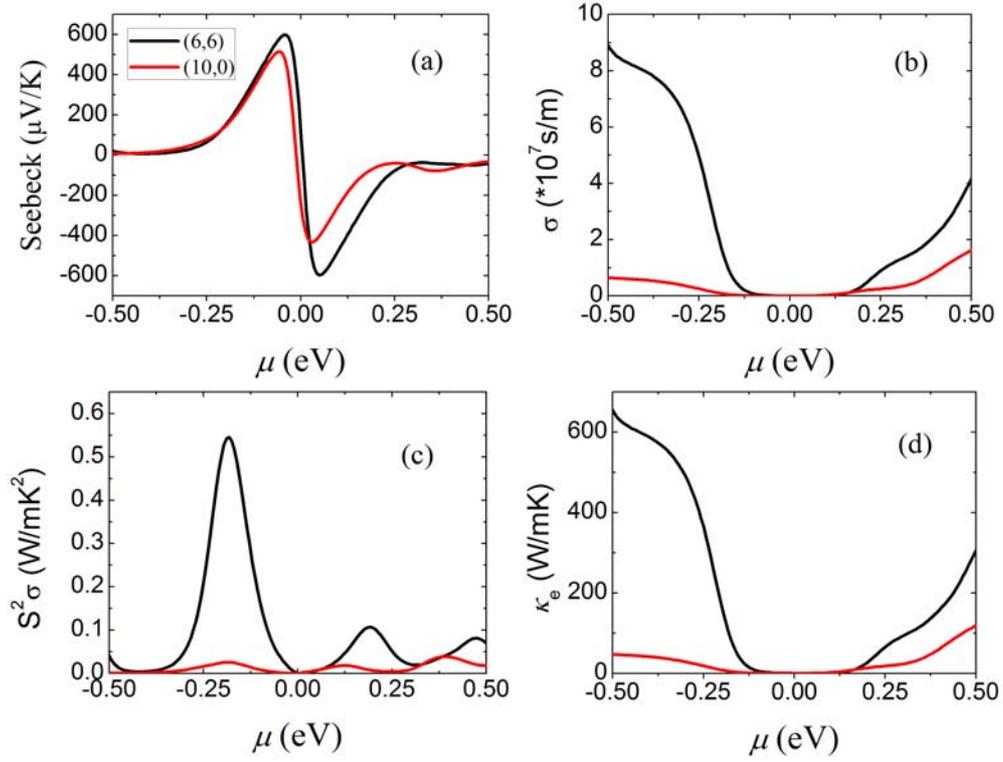

**Fig. 4** Calculated room temperature transport coefficients as a function of chemical potential for the SiGe (6, 6) and (10, 0) nanotubes: (a) Seebeck coefficient, (b) electrical conductivity, (c) power factor, and (d) electronic thermal conductivity.

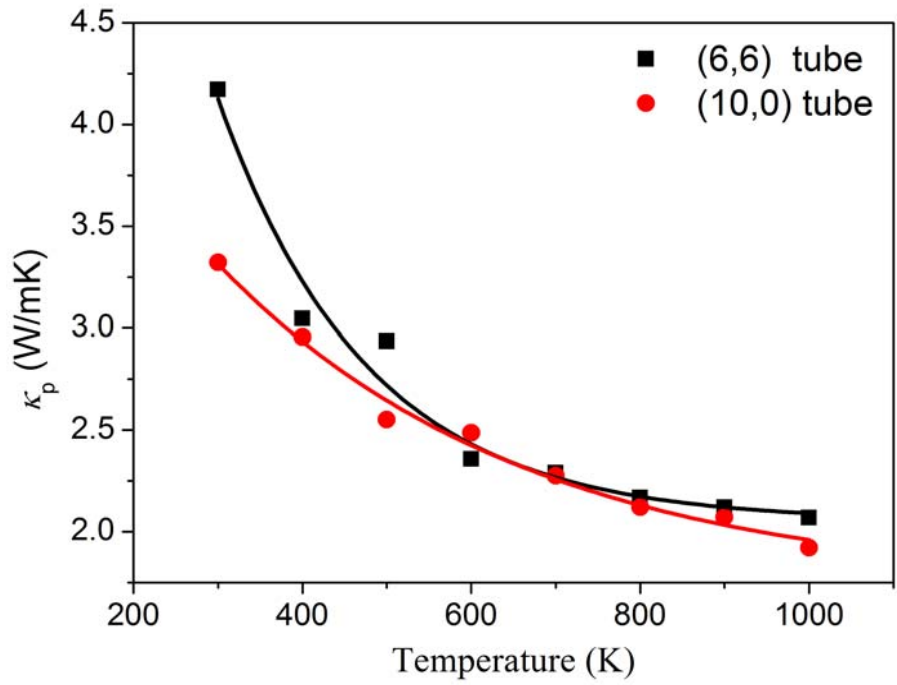

**Fig. 5** Calculated phonon-derived thermal conductivity of the SiGe (6, 6) and (10, 0) nanotubes as a function of temperature. The solid lines represent exponential fitting of the calculated results.

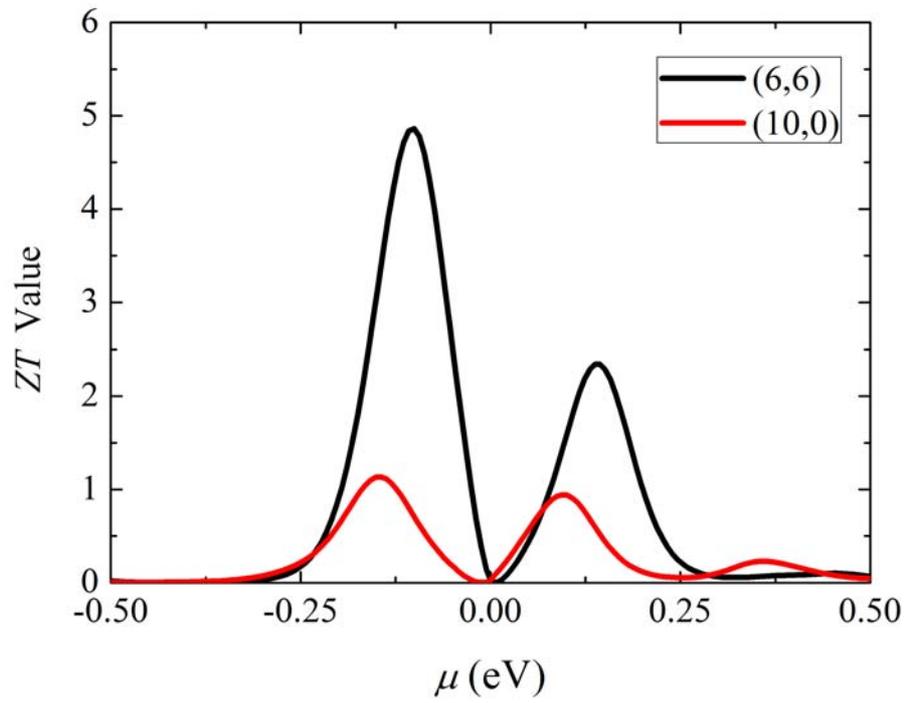

**Fig. 6** Calculated room temperature *ZT* values as a function of chemical potential for the SiGe (6, 6) and (10, 0) nanotubes.

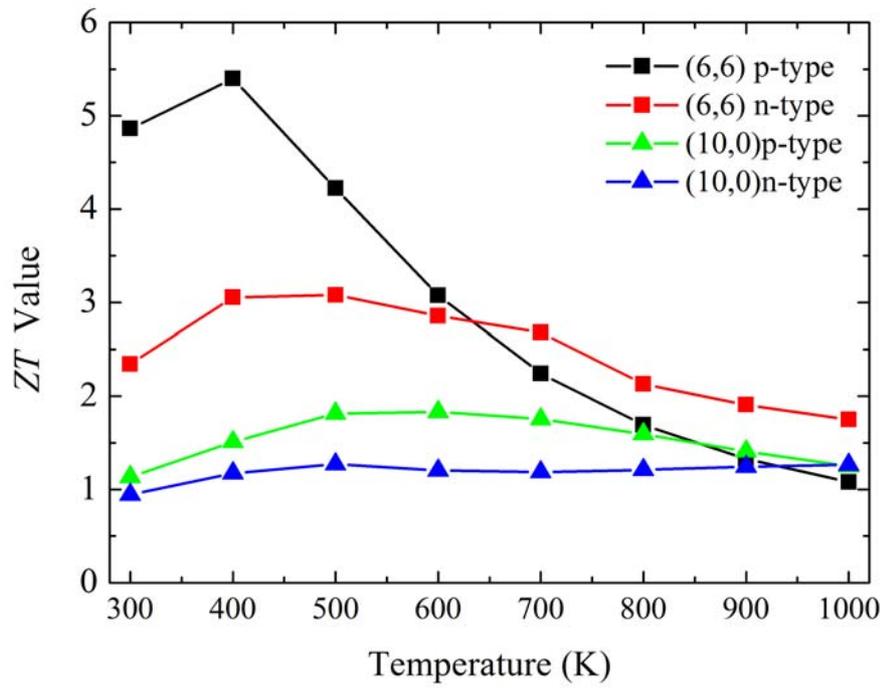

**Fig. 7** The temperature dependence of *ZT* values for the SiGe (6, 6) and (10, 0) nanotubes.